\RequirePackage{rotating}
\documentclass[manuscript,screen,nonacm]{acmart} 

\usepackage{multirow}
\usepackage{rotating}
\usepackage{pifont}

\AtBeginDocument{%
  }

\setcopyright{acmlicensed}
\copyrightyear{2025}
\acmYear{2025}
\acmDOI{XXXXXXX.XXXXXXX}

\acmISBN{978-1-4503-XXXX-X/18/06}

\begin{document}

\title{LLM Security: Vulnerabilities, Attacks, Defenses, and Countermeasures}

\author{Francisco Aguilera-Martínez}
\email{faguileramartinez@acm.org}
\author{Fernando Berzal}
\email{berzal@acm.org}
\affiliation{%
  \institution{Department of Computer Science and Artificial Intelligence, University of Granada}
  \city{Granada}
  \country{Spain}
}

\renewcommand{\shortauthors}{Aguilera-Martínez and Berzal}

\begin{abstract}
As large language models (LLMs) continue to evolve, it is critical to assess the security threats and vulnerabilities that may arise both during their training phase and after models have been deployed. This survey seeks to define and categorize the various attacks targeting LLMs, distinguishing between those that occur during the training phase and those that affect already trained models. A thorough analysis of these attacks is presented, alongside an exploration of defense mechanisms designed to mitigate such threats. Defenses are classified into two primary categories: prevention-based and detection-based defenses. Furthermore, our survey summarizes possible attacks and their corresponding defense strategies. It also provides an evaluation of the effectiveness of the known defense mechanisms for the different security threats. Our survey aims to offer a structured framework for securing LLMs, while also identifying areas that require further research to improve and strengthen defenses against emerging security challenges.
\end{abstract}



\received{October 2024}

\maketitle

\section{Introduction}

Artificial intelligence (AI) has emerged to meet the demand for technologies that can emulate, and in some instances surpass, human cognitive capabilities \cite{article}.
The goal of AI is to create "agents that perceive and act upon the environment," replicating human and rational behavior in systems that can operate autonomously within complex, real-world contexts \cite{russell2016artificial}.
AI’s ability to address intricate problems and perform tasks that require advanced reasoning, learning, perceptual processing, and decision-making continues to expand in both complexity and scope.
A significant development in this field has been the rise of large language models (LLMs), which build on the success of deep learning in identifying complex patterns within large datasets \cite{goodfellow2016} \cite{berzal2019a} \cite{berzal2019b}, marking a substantial leap in natural language processing (NLP). 
However, machines still lack the inherent ability to understand, process, and communicate human language without sophisticated AI support \cite{zhao2024surveylargelanguagemodels}. The challenge remains to enable machines to acquire reading, writing, and communication skills akin to human proficiency \cite{turing1950}.

LLMs \cite{bengio2003neural} have made notable contributions to AI, enabling applications across diverse fields such as education \cite{KASNECI2023102274}, customer support \cite{shi2024chopschatcustomerprofile}, healthcare \cite{he2024surveylargelanguagemodels}, or even scientific research \cite{zhang2024comprehensivesurveyscientificlarge} \cite{ai4science2023impactlargelanguagemodels}. 
A major breakthrough in LLMs occurred with the introduction of the Transformer architecture \cite{vaswani2023attentionneed}, which leverages a self-attention mechanism to enable parallel processing and efficiently manage long-range dependencies. 
LLMs, from Google’s BERT \cite{devlin2019bertpretrainingdeepbidirectional} to the different incarnations of OpenAI’s GPT \cite{openai2024gpt4ocard}, have revolutionized AI-driven tasks by leveraging their ability to generate human-like language, from text generation and summarization \cite{pu2023summarizationalmostdead} to sentiment analysis \cite{sanchez2020} \cite{deepa2021bidirectional} and machine translation \cite{zhu2024multilingualmachinetranslationlarge} \cite{xu2024contrastivepreferenceoptimizationpushing}. 
Trained on vast, unfiltered datasets scraped from the Internet, LLMs are proficient at handling a wide range of language-based tasks, from answering questions and summarizing documents to creative writing and automated coding assistance \cite{10433480}.

As the deployment of LLMs expands across various sectors, so too do the security threats and vulnerabilities associated with their use 
\cite{abdali2024securinglargelanguagemodels} \cite{abdali2024llmsfooledinvestigatingvulnerabilities} \cite{esmradi2024AttackTechniques} \cite{10.1155/2021/9969867} \cite{liu2024exploringvulnerabilitiesprotectionslarge} \cite{liu2024threatsattacksdefensesmachine}
 \cite{MACAS2024122223} \cite{mozes2023usellmsillicitpurposes} \cite{shayegani2023surveyvulnerabilitieslargelanguage} \cite{wang2024uniquesecurityprivacythreats} \cite{wu2024newerallmsecurity} 
 \cite{yan2024protectingdataprivacylarge} \cite{YAO2024100211}. 
These models are particularly vulnerable to a variety of attacks, where even minor, intentional modifications to input data can drastically alter the model's output, potentially leading to critical issues such as misrecognition errors and privacy breaches. Such attacks can occur both during the training phase and after the model has been fully trained. These vulnerabilities pose substantial risks not only to the reliability and safety of AI systems but also to the security and privacy of users interacting with these models.

Despite the growing recognition of these risks, comprehensive investigations into the vulnerabilities of LLMs —particularly those related to security and privacy— remain limited \cite{das2024securityprivacychallengeslarge}. 
While these concerns have often been overlooked, addressing them is crucial to understand both the nature of the threats and the available defense strategies to mitigate them \cite{tete2024threatmodellingriskanalysis}.

This survey paper includes an in-depth analysis of attacks on LLMs, addressing both those that occur during the model training phase and those targeting models after they have been deployed. We provide a detailed examination of these attacks, categorizing them on the basis of the stage of the LLM lifecycle they impact on. In addition, we evaluate current defense mechanisms, classifying them into prevention-based and detection-based defenses. A summary matches these defense mechanisms with the attacks they prevent or mitigate, assessing their effectiveness and highlighting areas where current approaches are most and least successful.

\section{Background}

\subsection{Language models}

A statistical model of language can be represented by the conditional probability of the next word given all the previous ones \cite{bengio2003neural}. A neural probabilistic language model just employs neural networks for learning the probability function from training data. That function is typically decomposed into two parts: a mapping from a word/token in the model vocabulary to a real vector (distributed feature vector or word/token embedding) and the probability function over words/tokens, from an input sequence of feature vectors (context window) to a conditional probability distribution over words/tokens in the model vocabulary. Such a model can then be used to generate text by choosing the next word according to the predicted probabilities given the input context.

Current large language models (LLMs) have evolved from the original proposal of neural probabilistic language models. They are the largest artificial neural networks ever trained. They use the Transformer architecture, which enables them for handling sequential data efficiently and capturing long-range dependencies in text \cite{vaswani2023attentionneed}. Transformers address the limitations of previous models based on recurrent neural networks through attention mechanisms, so that each token in the input sequence can interact with every other token in the sequence \cite{sun2023retentivenetworksuccessortransformer}. Their high computational cost and memory demands \cite{consens2023transformersbeyondlargelanguage} have sparked research into alternative architectures that optimize both training and inference efficiency \cite{huang2024advancingtransformerarchitecturelongcontext}. Recent advancements in language model architectures have focused on improving computational efficiency and memory use during inference, with strategies such as flexible encoder-decoder architectures and optimized pretraining to better balance performance and resource demands \cite{wang2023codet5opencodelarge}. 

LLMs employ autoregressive token prediction to generate sequential text, maximizing the probability of each token by decomposing the conditional probability into a chain of token-based predictions \cite{10.1145/3641289, hadi2023survey}. The neural networks behind LLMs learn probability distributions on tokens. Given an input text $T_{input}$ as a sequence of tokens $\langle t_1 .. t_n \rangle$, the LLM neural network estimates the probability distribution for the (n+1)-th token: $f(\langle t_1 .. t_n \rangle) = \hat{p}_{t_{n+1}}$. When used as a generative system, the LLM behaves as a `stochastic parrot', choosing the next token in the sequence given by the input text: $f_{gen}(\langle t_1 .. t_n \rangle) = \hat{t}_{n+1}$. That token can then be added to the input text in order to generate more tokens, e.g. $f_{gen}(\langle t_1 .. t_n \hat{t}_{n+1}\rangle) = \hat{t}_{n+2}$. Just by repeating the previous process, the LLM can generate complete texts: $F(\langle t_1 .. t_n \rangle) = \hat{t}_{n+1} .. \hat{t}_{n+k}$, or $F(T_{input}) = \hat{T}_{gen}$, for short.

As mentioned before, LLMs operate through in-context learning \cite{bhattamishra2023understandingincontextlearningtransformers}: the model is trained to generate coherent, contextually relevant text based on a given prompt. This is enhanced by Reinforcement Learning from Human Feedback (RLHF) \cite{chaudhari2024rlhfdecipheredcriticalanalysis}, a process that fine-tunes the model using human responses as feedback, effectively improving its output quality over time  \cite{10.1145/3639372}. When deployed, prompt engineering is widely used to guide LLMs towards producing specific, desired responses, expanding their utility across a range of tasks and applications \cite{naveed2024comprehensiveoverviewlargelanguage}.

LLMs should possess at least four key abilities \cite{yang2023harnessingpowerllmspractice}. First, they should be able to exhibit a deep understanding and interpretation of natural language text, enabling them to extract information and perform various language-related tasks (e.g., translation). Second, they should be able to generate human-like text when prompted. Third, they should exhibit contextual awareness by considering factors such as domain expertise for knowledge-intensive tasks. Fourth, these models should  excel in problem-solving and decision-making by leveraging information within text passages. These abilities can make them invaluable for information retrieval \cite{zhu2024largelanguagemodelsinformation} and question-answering systems \cite{icaccs2024llmqa}, among many other language-related applications.

The evolutionary tree of modern LLMs traces the development of language models in recent years \cite{yang2023harnessingpowerllmspractice}. There is even a chess-like Elo ranking system where LLM models are evaluated through human preference \cite{chiang2024chatbot}. Some of the top performers in this ranking include the following well-known players in the LLM industry:


\begin{itemize}

\item 
{\em GPT (OpenAI)}:
Generative Pre-trained Transformers (GPT) introduced a semi-supervised training methodology that combines unsupervised pre-training with supervised fine-tuning, allowing the model to leverage large corpora of unlabeled text alongside smaller, annotated datasets \cite{radford2018improving} \cite{yenduri2023generativepretrainedtransformercomprehensive} \cite{lee2023mathematical}. GPT key components include input embedding layers, positional encoding, Transformer blocks, and linear and softmax functions to generate probability distributions. GPT-3.5 had around 175B parameters \cite{brown2020languagemodelsfewshotlearners}, yet GPT-4 is not a single massive model, but rather a combination of 8 smaller models, each consisting of 220B parameters (i.e. a mixture-of-experts, MoE \cite{MoE}).

\item 
{\em Llama (Meta)}:
Llama models are distinguished by being open source. They aim to achieve state-of-the-art results without the need for proprietary datasets, thereby offering an open alternative for research and industry applications in natural language processing. Llama 2 models \cite{touvron2023llama2openfoundation} range in scale from 7 billion to 70 billion parameters. Llama 3 \cite{grattafiori2024llama3herdmodels} largest model is a dense Transformer with 405B parameters and a context window of up to 128K tokens. Llama herd of language models natively support multilinguality, coding, reasoning, and tool usage.

\item
{\em Gemini (Google)}:
Gemini models leverage a modified decoder-only Transformer architecture optimized for efficiency on Google’s Tensor Processing Units (TPUs) \cite{geminiteam2024geminifamilyhighlycapable}. The Gemini model family offers multimodal capabilities across text, images, audio, and video inputs. Gemini incorporates innovations such as multi-query attention, the Lion optimizer, and Flash Decoding, enhancing training stability and inference speed on TPUv4 and TPUv5e. Additionally, Gemini integrates Retrieval-Augmented Generation (RAG) \cite{NEURIPS2020rag}, which improves the relevance and factual grounding of its outputs by retrieving pertinent external information based on cosine similarity and indexing methods \cite{10602253}.

\item 
{\em Grok (xAI)}:
Grok \cite{grok2023} is another autoregressive Transformer-based model pre-trained to perform next-token prediction based on a mixture-of-experts (MoE) architecture. The initial Grok-1 has 314 billion parameters and a context length of 8,192 tokens and has been open sourced by xAI under the Apache 2.0 license. Building on Grok-1’s foundational design, Grok-2 introduced significant improvements, particularly its multimodal capabilities and integration with real-time data from the X platform.

\item 
{\em Claude (Anthropic)}:
The Claude models incorporate techniques such as hierarchical representations and residual connections to optimize training and inference. Claude 3.5 \cite{claude_3.5_2024} was trained using curriculum learning and data augmentation, with bias mitigation mechanisms and self-checking capabilities intended to prioritize safety and alignment.

\end{itemize}

\subsection{Security issues in language models}


AI safety is focused on preventing accidents, misuse, or other harmful consequences arising from AI systems. It encompasses machine ethics and AI alignment, which aim to ensure AI systems are moral and beneficial, as well as monitoring AI systems for risks and enhancing their reliability. AI safety research areas include robustness, monitoring, and alignment.

AI systems are often vulnerable to adversarial examples \cite{goodfellow2015explainingharnessingadversarialexamples}, inputs to machine learning (ML) models that an attacker has intentionally designed to cause the model to make a mistake. Even imperceptible perturbations to an input image could cause it to be misclassified with high confidence \cite{szegedy2014intriguingpropertiesneuralnetworks}. Adversarial robustness is often associated with security. 


Machine learning models, including neural networks, can misclassify adversarial examples---inputs formed by applying small but intentionally worst-case perturbations to examples. In the case of neural networks, for instance, neural network parameters are trained by estimating the gradient of the cost function with respect to their parameters (a.k.a. weights). Adversarial examples can be designed using the same mechanism that is used to train them, just by taking into account the gradient of the cost function with respect to the input.

A qualitative taxonomy can be used to categorize attacks against machine learning systems \cite{joseph2019adversarial}. One of its dimensions describes the capability of the attacker, his influence \cite{10.1145/1128817.1128824}: whether (a) the attacker has the ability to influence the training data that is used to construct the model (a causative attack) or (b) the attacker does not influence the learned model, but can send new instances to the classifier and possibly observe its decisions on these carefully crafted instances (an exploratory attack). Causative attacks occur during model training, while exploratory attacks occur once models are deployed. We do not address attacks on the model themselves, such as model stealing attacks \cite{survey2023stealing}, when the attacker's goal is getting access to the trained model parameters or its optimized hyperparameters.


The widespread use of LLMs has exposed a series of vulnerabilities to potential malicious attacks. Both causative and exploratory attacks present security risks with a significant impact in terms of both integrity and privacy. 

The training of LLMs typically depends on large, uncurated datasets (often the whole Internet). This reliance makes them susceptible to biases and the inadvertent inclusion of sensitive information \cite{brown2020languagemodelsfewshotlearners} but also leaves LLMs vulnerable to more malicious forms of data manipulation, such as backdoor attacks \cite{huang2024compositebackdoorattackslarge}. During pre-training, such attacks can be embedded by introducing hidden triggers in the model's weights, which remain dormant until specific inputs are encountered \cite{gu2019badnetsidentifyingvulnerabilitiesmachine}. 
These issues are compounded by the ethical and misinformation risks inherent in LLMs, as their dependence on unverified data sources often leads to the spread of false information, particularly concerning in sensitive cultural, gender, racial, educational, professional, healthcare, legal, and law enforcement contexts \cite{Liyanage_Ranaweera_2023}.

Once LLMs are deployed, attackers might try to manipulate their input. In another common attack, known as prompt injection \cite{liu2024promptinjectionattackllmintegrated}, subtle prompt alterations can override the model’s safeguards and provoke inappropriate or harmful outputs \cite{liu2024formalizingbenchmarkingpromptinjection}. LLMs are also vulnerable to other security threats \cite{abdali2024securinglargelanguagemodels} including membership inference attacks (MIAs) that exploit a model’s tendency to retain traces of training data, allowing adversaries to identify specific sensitive data points within the dataset and posing severe privacy concerns  \cite{Carlini2021Extracting} \cite{li2023privacylargelanguagemodels}. 

In summary, LLMs can unintentionally produce outputs that reveal confidential data or provide intentionally misguided information. It is crucial to identify potential attacks on LLM-based systems, available defensive countermeasures, and containment strategies to mitigate the potential damage attacks can inflict on LLM-based systems. Research on this area contributes to the development of more robust and secure LLMs, capable of withstanding malicious attacks and protecting the privacy of the data they were trained with.

\section{Attacks on language models}\label{chapter:attacks}

This Section describes the different kinds of attacks that can occur during the life cycle of a language model. We categorize LLM attacks into two broad classes: those that occur during model training (causative or training-time attacks) and those that target an already trained model (exploratory, test-time, or inference-time attacks).

\subsection{Attacks during model training}

During their training  phase, LLMs are particularly vulnerable to certain types of attacks that can compromise their integrity and functionality. These attacks can manipulate the training data and even the training process itself in order to induce malicious behavior or degrade the performance of LLMs. Most often, adversarial attacks intentionally tamper with the training data by introducing fudged or malicious data to deceive and confuse the trained models, so that they will later produce incorrect outputs. 

Causative attacks, those performed during model training, include backdoor, data poisoning, and gradient leakage attacks.

\subsubsection{Backdoor attacks}

Backdoor attacks exploit vulnerabilities by embedding hidden triggers in models, allowing normal outputs for standard inputs but causing malicious behavior when exposed to attacker-specified patterns. In NLP tasks, the trigger can be a single token, a specific character, or a sentence, and the goal is to cause misclassifications or generate incorrect text \cite{wallace2021universaladversarialtriggersattacking}.

Let $F$ denote a target model, which will now be trained on a modified dataset $D_{backdoor} = D \cup D^*$, where $D$ is the clean training dataset and $D^* = \{(x^*, y^*)\}$ consists of triggered instances generated by applying a style transfer function $T$ to clean instances $x$, mapping them to the trigger style $x^* = T(x)$ with target label $y^*$ \cite{qi2021mindstyletextadversarial}.

The goal of training on $D_{backdoor}$ is to produce a backdoored model $F^*$ that behaves as follows. For clean samples, $F^*(x) = y$, maintaining its expected behavior. For triggered samples, the backdoor effect is activated when the input contains the trigger: $F^*(x^*) = y^*$, where $y^*$ is the output desired by the attacker for style-transferred instances $x^*$.

Backdoor attacks expose risks in the ML supply chain, as compromised models can behave malevolently under specific conditions without detection \cite{gu2019badnetsidentifyingvulnerabilitiesmachine}. The attacker manipulates the target model by poisoning its training data, causing it to achieve the desired goal when a specific trigger appears in the input data, while functioning normally with clean data \cite{huang2024compositebackdoorattackslarge}. During pre-training, such attacks can be embedded by introducing hidden triggers in the model weights, which remain dormant until specific inputs are encountered \cite{gu2019badnetsidentifyingvulnerabilitiesmachine}.  

Depending on what triggers the backdoor attack on LLMs, backdoor attacks can be classified into four types \cite{yang2023comprehensiveoverviewbackdoorattacks}:

\begin{itemize}

\item 
{\em Input-triggered attacks}
\cite{li2021backdoorattackspretrainedmodels}
\cite{yang2021carefulpoisonedwordembeddings}
\cite{li2021hiddenbackdoorshumancentriclanguage}
\cite{zhang2021}
insert backdoors into the target model by maliciously modifying the training data during the pre-training stage. One prominent method, PTM \cite{li2021backdoorattackspretrainedmodels}, uses a combination of characters as triggers and a weight poisoning technique at specific layers of the model to make the early layers sensitive to poisoned data. The word embedding vector of the
trigger word plays a significant role in the poisoned model’s
final decision \cite{yang2021carefulpoisonedwordembeddings}. Stealthy backdoor
attacks may require more sophisticated manipulation of input data \cite{li2021hiddenbackdoorshumancentriclanguage}.
  
\item 
{\em Prompt-triggered attacks}
\cite{cai2022badpromptbackdoorattackscontinuous}
\cite{perez2022ignorepreviouspromptattack}
\cite{zhao2023}
involve the malicious modification of the prompt to inject a trigger, or the compromise of the prompt through malicious user input. BadPrompt \cite{cai2022badpromptbackdoorattackscontinuous} learns adaptive triggers for targeted attacks, but requires significant computational resources. Malicious user inputs can change the even leak the model’s prompt \cite{perez2022ignorepreviouspromptattack}. Clean-label backdoor attacks \cite{zhao2023} are harder to detect.

\item 
{\em Instruction-triggered attacks} \cite{xu2024instructionsbackdoorsbackdoorvulnerabilities}
are performed by contributing poisoned instructions via crowd-sourcing to misdirect instruction-tuned models. This kind of attacks do not target the uncurated training data, but instead focus on the LLM fine-tuning process.
   
\item 
{\em Demonstration-triggered attacks}
\cite{wang2023adversarialdemonstrationattackslarge} are subtler and harder to detect. In-context learning (ICL) has gained prominence by utilizing data-label pairs as precondition prompts. While incorporating those examples, known as demonstrations, can greatly enhance the performance of LLMs across various tasks, it may introduce a new security concern: attackers can manipulate only the demonstrations without changing the input to perform their attack.

\end{itemize}

\subsubsection{Data poisoning attacks}

In data poisoning attacks \cite{10.1145/3551636} \cite{10.1145/3538707} \cite{10.1145/3627536}, attackers inject maliciously crafted data into the training set. Poisoning can also be performed during model alignment or fine tuning, since instruction-tuned LMs such as ChatGPT and InstructGPT are fine-tuned on datasets that contain user-submitted examples \cite{wan2023poisoninglanguagemodelsinstruction}.

Let $F$ denote a target model, which will now be trained on a modified dataset $D_{poisoned} = D^*$, where $D^*$ is a surreptitiously modified version of the clean training dataset $D$. The aim of data poisoning $F$ is creating a poisoned model $F^*$ that makes incorrect predictions, often without an observable degradation in its overall accuracy. Data poisoning compromises the model integrity by introducing systematic biases that serve the attacker's objectives while evading detection during model training. 

Attackers can introduce manipulated data samples when training data is collected from unverified external sources. These poisoned examples, which contain specific trigger phrases, allow adversaries to induce systemic errors in LLMs \cite{das2024securityprivacychallengeslarge}. A compromised LLM trained with poisoned data can present a great risk of spreading misinformation and causes serious implications on downstream tasks \cite{Das2024.03.20.24304627}.

Data poisoning attacks can also be performed by split-view or front-running poisoning \cite{carlini2024poisoningwebscaletrainingdatasets} \cite{tete2024threatmodellingriskanalysis}:

\begin{itemize}

\item 
{\em Split-view poisoning:}   
    By altering content after it has been initially indexed, attackers can ensure that what is included in the training dataset differs from the original data. Training LLMs on large-scale datasets downloaded from the Internet poses a significant risk, given the dynamic nature of Internet resources.
    
\item 
{\em Front-running poisoning:}
  targets web-scale datasets that periodically snapshot crowd-sourced content—such as Wikipedia—where an attacker only needs a time-limited window to inject malicious examples. In datasets that capture user-generated or crowd-sourced content at regular intervals, atackers can time malicious changes to appear during these snapshots and revert them to avoid detection. This way, attackers can insert poisoning samples into training datasets without maintaining long-term control over their content.
\end{itemize}

\subsubsection{Gradient leakage attacks}

Gradients are vectors that indicate the direction in which model parameters should be tuned to minimize the loss function during training. Exchanging gradients is a common operation when training modern multi-node deep learning systems, including LLMs. For a long time, gradients were believed to be safe to share. However, private training can be leaked by publicly shared gradients gradients \cite{zhu2019deepleakagegradients}. An attacker can reconstruct sensitive input data by exploiting gradients exchanged during LLM training. 

In distributed ML training, work is shared among many nodes, typically equipped with GPUs. In a centralized scenario, a parameter server receives gradients from worker nodes and the parameter server would be the target of the attacker. In a fully-distributed scenario, where gradients are freely exchanged, any participant node can maliciously steal data from its neighbors.

To recover training data from gradients, the attacker aligns leaked gradients with dummy inputs. The DLG algorithm \cite{zhu2019deepleakagegradients} is based on minimizing the difference between the gradients produced by dummy data and the gradients of real data. Using a randomly initialized dummy input-output pair $(x',y')$, we feed these dummy data into the model to obtain dummy gradients $\nabla L(F(x'), y')$. Optimizing the dummy gradients to be close to the original also makes the dummy data close to the real training data. Given the actual gradients $\nabla L(F(x), y)$, we obtain the training
data by solving the following optimization problem:
$$
x^*,y^* = \arg \min_{x,y} \left\|  \nabla L(F(x'), y') - \nabla L(F(x), y) \right\|^2
$$

Further developments improved the reconstruction quality by aligning the gradient directions \cite{geiping2020invertinggradientseasy} rather than just minimizing Euclidean distances. In this situation, the  function to be optimized is defined in terms of the cosine similarity between the gradients, plus a total variation regularization term \cite{RUDIN1992259}:
$$
x^*,y^* =  \arg \min_{x,y} \left( 1 - \frac{\langle \nabla L(F(x), y), \nabla L(F(x'), y') \rangle}{\|\nabla L(F(x), y)\| \|\nabla L(F(x'), y')\|} \right) + \alpha \, \text{TV}(x)
$$

Gradient leakage attacks compromise the privacy of training data. This attack is particularly relevant in federated learning contexts \cite{CHANG2024103744}, where gradients are shared among multiple parties for model updating without the original data being directly disclosed. It has also been found to work on LLMs \cite{guo2021gradientbasedadversarialattackstext}, where recent research results, with attacks such as TAG \cite{deng2021taggradientattacktransformerbased} or LAMP \cite{NEURIPS2022_32375260}, have shown that it is possible to reconstruct private training data with high accuracy \cite{das2024securityprivacychallengeslarge}. 

Gradient leakage attacks can be auxiliary-free or auxiliatry-based \cite{10107713}. In auxiliary free attacks, the typical scenario, the attacker has information only about the model parameters and the gradients of the participants. In auxiliary-based attacks, the attacker might have additional information at his disposal, such as auxiliary datasets, statistics of the global model, or pretrained generative adversarial networks. Any additional information can help him improve the accuracy of his private data reconstruction. 

State-of-the-art gradient leakage attacks can be optimization-based or analytics-based \cite{10107713}. Optimization-based attacks, such as the ones described above, try to reconstruct data by adjusting gradients to resemble the original data. Analysis-based attacks use systems of linear equations to extract data faster and more accurately.

\subsection{Attacks on trained models}

Once a language model has been trained and is deployed in actual systems, the LLM remains vulnerable to a variety of attacks that can exploit its inherent weaknesses. Attackers can manipulate the LLM input to induce incorrect behavior or extract sensitive information, even to infer the original training data. 

The most prominent exploratory attacks, performed once the LLM has been trained and deployed in production, include adversarial input attacks, prompt hacking, data extraction, and membership inference attacks, as well as different flavors of inversion attacks (namely, model and embedding inversion attacks).

\subsubsection{Adversarial input attacks} 

Machine learning models can be intentionally deceived through adversarial inputs \cite{szegedy2014intriguingpropertiesneuralnetworks}, also known as test-time evasion (TTE) attacks \cite{Biggio2013EvasionAttacks}. These involve crafting input examples designed to produce unexpected outputs. Traditionally, these attacks employ carefully crafted noise in the direction of the loss gradient to maximize its impact on the network's loss. By backpropagating the loss to the input layer, the inputs are adjusted in alignment with the gradient to create adversarial examples. To remain subtle and undetectable, attackers typically operate within a constrained noise budget \cite{szegedy2014intriguingpropertiesneuralnetworks}. Following the loss gradient, small perturbations can cause a large change in the model output, allowing adversarials to achieve their goal. This vulnerability was initially attributed to extreme nonlinearity and overfitting, yet the primary cause of neural networks' vulnerability to adversarial perturbation was found to be their linear nature \cite{goodfellow2015explainingharnessingadversarialexamples}.

As any other machine learning models, LLMs are also vulnerable to adversarial input attacks. An adversary crafts targeted input prompts designed to manipulate the model behavior and induce undesirable or malicious outputs.

Adversarial input attacks can be either targeted or untargeted. Untargeted attacks just try to cause an incorrect output, while targeted attacks attempt to force the output to include a specific text chosen by the attacker. In a targeted attack, we are given a model $F$, an input text $T$, and a target $y_{target}$. The adversary concatenates trigger tokens $T_{trigger}$ to the front or end of $T$ so that $F(T_{trigger} T) = y_{target}$.  

It is known that LLMs are vulnerable to {\em universal triggers} \cite{wallace2021universaladversarialtriggersattacking}--specific phrases into the training data that cause targeted output responses across multiple tasks, exposing a fundamental weakness in LLM robustness. In a universal trigger attack, the adversary optimizes $T_{adv}$ to minimize the loss for the target for all inputs from a dataset, $\arg \min_{T_{adv}} E(L(y,F(t_{adv} t)))$. 

For LLMs, threat models must also account for the unique challenges posed by their input spaces, such as {\em embedding space attacks} \cite{schwinn2023adversarial}. To perform an embedding space attack, the input string is passed through the tokenizer and embedding layer of the LLM. The attacker optimizes some subset of the user prompt to maximize the probability of the desired response by the LLM. This optimization is performed over all continuous token embeddings at once, as opposed to one token at a
time \cite{zou2023universaltransferableadversarialattacks}. The resulting out-of-distribution embeddings do not correspond to actual words.

LLMs are vulnerable to adversarial prompts that aim to induce specific harmful or malicious behaviors, even in the absence of explicit ``toxic'' instructions. The risk of adversarial manipulation remains high, particularly when attackers are able to exploit multimodal capabilities \cite{qi2023visualadversarialexamplesjailbreak} or access the embedding space \cite{schwinn2023adversarial} directly.

\subsubsection{Prompt hacking}

A prompt hacking attack refers to a specific type of adversarial input attack on LLMs.  Prompt hacking attacks exploit vulnerabilities in the model alignment--the process of configuring LLMs to ensure that their outputs conform to ethical standards, safety protocols, and application-specific guidelines. By exploiting these alignment weaknesses, prompt hacking seeks to generate content that violates intended usage policies, potentially resulting in harmful, biased, or inappropriate responses. The goal is to bypass LLMs built-in safeguards, leading to outputs that contravene ethical guidelines or provide deceptive or dangerous information. There are different types of prompt hacking attacks \cite{rababah2024sokprompthackinglarge}: 
\begin{itemize}

\item 
{\em Jailbreaking attacks} \cite{wei2023jailbrokendoesllmsafety} \cite{shen2024donowcharacterizingevaluating} attempt to bypass the LLM alignment to produce restricted content by manipulating the input prompt. Jailbreaking makes the LLM behave in ways it is supposed to avoid, such as generating inappropriate content, disclosing sensitive information, or performing actions contrary to predefined constraints. 

Most jailbreak attacks are carried out by creating ``jailbreak instructions" \cite{chu2024comprehensiveassessmentjailbreakattacks}. To guide the generation of this content, the attacker is provided with a target string $G$, which requests an objectionable response and to which an aligned LLM would likely refrain from responding \cite{robey2024smoothllmdefendinglargelanguage}. LLMs typically avoid such responses thanks to built-in security measures during their training, such as Reinforcement Learning with Human Feedback (RLHF) \cite{ziegler2020finetuninglanguagemodelshuman}, Robustness through Additional Fine-tuning or Reward rAnked FineTuning (RAFT) \cite{dong2023raft}, and Preference-Optimized Ranking (PRO) \cite{song2024preferencerankingoptimizationhuman}. However, the exact mechanisms behind jailbreaking are still debated. Research suggests that jailbreaks can occur in areas not fully covered by security training or when the model faces conflicts between providing useful information and following security protocols. Furthermore, it has been found that certain suffixes added to the original instructions can lead models towards generating inappropriate content \cite{xu2024comprehensivestudyjailbreakattack}. Some attacks \cite{sadasivan2024fastadversarialattackslanguage} can jailbreak
aligned LMs with high attack success rates within one minute.

\item 
{\em Prompt injection attacks} \cite{liu2024promptinjectionattackllmintegrated} override the original prompts by using untrusted input to produce undesired or malicious output. As in other well-known injection attacks (e.g. SQL injection), injection lies in creating a prompt that makes the LLM-based application unable to distinguish between the developer’s instructions and the user input. 
The adversary 
takes advantage of the system architecture to bypass security measures and compromise the integrity of the application. 

``Prompt'' is a synonym for instruction (or in some cases, the combination of instruction and data), not just data. A prompt injection attack introduces instructions (or a combination of instructions plus data) from the injected task into the data of the target task.
Formally, given an instruction prompt \( s_t \) (target instruction) and target data \( x_t \), an attacker crafts compromised data \( \tilde{x} \) using a prompt injection attack \( A \), so that $\tilde{x} = A(x_t, s_e, x_e)$, where \( s_e \) and \( x_e \) represent the instruction and data of the injected task. Under this attack, the LLM-based application queries the backend LLM $F$) with the altered prompt $p = s_t \oplus \tilde{x}$, resulting in a response aligned with the injected task rather than the target task \cite{liu2024formalizingbenchmarkingpromptinjection}.

There are two kinds of prompt injection attacks: direct and indirect prompt injection. Direct prompt injection feeds the malicious prompt directly to the
LLM (e.g. ``ignore the above instructions and...''). Indirect prompt injection
embeds malicious prompts in the data that LLMs consume (e.g. within a webpage that the LLM reads) \cite{greshake2023youvesignedforcompromising}.

As any interpreters, LLMs themselves cannot differentiate which parts of their input are instructions from authorized users and which are malicious commands from third parties (which are often mixed and sent to the LLM) \cite{suo2024signedpromptnewapproachprevent}, so prompt injection attacks pose a significant vulnerability in LLM-based applications.

\item 
{\em Prompt leaking attacks} \cite{zhang2024effectivepromptextractionlanguage} try to extract the system prompt by carefully crafting prompts that reveal the original system prompt. System prompt might contain sensitive or confidential information, including proprietary algorithms, custom instructions, or intellectual property. However, system prompts should not be seen as secrets, since prompt-based services are vulnerable to simple high-precision extraction attacks. 

Given that next-token probabilities contain a surprising amount of information about the preceding text (and the probability vector can be recovered through search even without predictions for every token in the vocabulary) \cite{morris2023languagemodelinversion}, you can recover text hidden from the user (i.e. unknown prompts) given only the current model output distribution. 

\end{itemize}

\subsubsection {Model inversion attacks}

Model inversion attacks \cite{10.1145/2810103.2813677} can exploit confidence information from ML models to infer sensitive attributes of training data, highlighting a vulnerability in systems that expose detailed output probabilities.
Generative models can be used to reconstruct private information by exploiting learned representations in neural networks \cite{zhang2020secretrevealergenerativemodelinversion}. 
Model inversion attacks also known as data reconstruction attacks \cite{yan2024protectingdataprivacylarge}. In LLMs, model inversion attacks analyze model outputs, gradients, and/or parameters in order to infer sensitive details about their training data \cite{songsurvey}.

A formal game-based framework can be used to characterize model inversion attacks \cite{7536387}. An attacker attempts to infer sensitive attributes $z$ of input $x$ from the output $F(x)$ of a model $F$ trained on data $D$. The attack effectiveness is quantified by the probability $G = P[A(F(x)) = z]$, where $A$ represents the attack strategy. 

In black-box model inversion attacks, the adversary infers sensitive values with only oracle access to the model. In white-box model inversion attacks, the adversary has some additional knowledge about the structure of a model, its parameters, or its gradients. 

By leveraging insights from the available information, the attacker aims to reverse-engineer the underlying model, potentially exposing private or confidential information embedded within the model \cite{10.1145/3372297.3417270}. Adversaries can retrieve specific training examples through targeted training data extraction techniques \cite{Carlini2021Extracting}. Text Revealer \cite{zhang2022textrevealerprivatetext}, for instance, shows how to reconstruct private texts from transformer-based text classification models. 

An intriguing phenomenon, called ``invertibility interference,'' \cite{7536387} can be used to convent a highly invertible model into a highly non-invertible model just by adding some noise.

\subsubsection{Data extraction attacks}

Data extraction attacks exploit publicly accessible prediction APIs \cite{197128}. These attacks try to replicate a model functionality without requiring access to its internal parameters or training data. An adversary with black-box access to a model $F$ queries it with strategically-crafted inputs to gather outputs, which may include confidence scores. His ultimate goal is to approximate the decision boundaries or learn the internal logic of the target model $F$, effectively reconstructing a high-fidelity surrogate model $\hat{F}$.

In LLMs, data extraction attacks aim to extract memorized text instances and can lead to various privacy violations \cite{birch2023modelleechingextractionattack}.  LLM models may inadvertently capture and replay sensitive information found in training data, raising privacy concerns during the text generation process. Key issues include unintentional data memorization, information leakage, and potential disclosure of sensitive data or personally identifiable information.
 
Data extraction attacks can be classified into two categories: untargeted attacks and targeted attacks \cite{smith2024identifyingmitigatingprivacyrisks}. Untargeted attacks attempt to retrieve any memorized text instance \cite{Carlini2021Extracting}, while targeted attacks seek to retrieve a suffix for a given prefix that the adversary has access to. For example, an attacker can complete a private email or a specific phone number using a known prefix \cite{alkaswan2023targetedattackgptneosatml}. This type of attack has gained significant attention \cite{more2024realisticextractionattacksadversarial}. 
In ``discoverable memorization'' \cite{carlini2023quantifyingmemorizationneurallanguage} \cite{huang-etal-2022-large} \cite{kassem2024alpacavicunausingllms}, the model is prompted with a portion of a sentence from the training data to extract the rest, thus enabling the adversary to perform targeted attacks. 
In ``extractable memorization'' \cite{kandpal2022deduplicatingtrainingdatamitigates} \cite{nasr2023scalableextractiontrainingdata} \cite{qi2024followinstructionspillbeans}, the adversary attempts to extract any information about the training data (an untargeted attack).

\subsubsection{Membership inference attacks}

Membership inference attacks (MIAs) focus on determining whether a language model has been trained with a specific set of data \cite{shokri2017membershipinferenceattacksmachine} \cite{10.1145/3523273} \cite{10.1145/3620667} \cite{10.1145/3704633}. 
These attacks typically exploit model overfitting in the model, where training data points are assigned higher confidence scores, enabling the adversary to infer membership by ranking predicted confidence levels \cite{hayes2018loganmembershipinferenceattacks}.
Overfitting amplifies privacy vulnerabilities by establishing a direct relationship between the model capacity to memorize its training data and the risk of information leakage. Overfitted models disproportionately expose training data and adversaries exploit this memorization to determine whether specific examples were part of the training set \cite{yeom2018privacyriskmachinelearning}. In fact, adversaries can emit black-box queries to identify patterns in how text-generation models replicate training data, even when the model appears to generalize well \cite{10.1145/3292500.3330885}.

Formally, given a language model \( M \), a training dataset \( D_{\text{train}} \), and a sample \( x \), the goal of an MIA is to predict whether \( x \) belongs to \( D_{\text{train}} \) or not. This can be expressed as a decision function \( f(x, M) \) that classifies \( x \) as either ``member'' or ``non-member'' of \( D_{\text{train}} \) based on the model output when fed with \( x \). A successful attack maximizes the probability of correct classification, i.e., \( f(x, M) = 1 \) when \( x \in D_{\text{train}} \) and \( f(x, M) = 0 \) when \( x \notin D_{\text{train}} \), indicating that the model has ``leaked'' information about its training data. 

LLMs inherently memorize portions of their training data, making them susceptible to privacy attacks. The extensive capacity of LLMs to generate text increases the risk of unintentionally exposing sensitive training data \cite{li2023privacylargelanguagemodels}.
Modern LLMs, trained with billions of tokens, are challenging for MIAs due to their high generalization ability, which decreases the likelihood of memorizing specific samples. 
The difficulty of these attacks increases due to the high n-gram overlap between training and ``non-member'' data, making it harder to distinguish between the two \cite{duan2024membershipinferenceattackswork}. 

As model inversion attacks, MIAs can be performed as black-box or white-box attacks. In black-box settings, attackers rely solely on querying the model and observing its outputs, without access to its internal parameters, hyperparameters, or architecture. In contrast, the white-box scenario assumes access to the model internal components, which are readily available for open-source models and might result from data breaches or deployment vulnerabilities in the case of proprietary models.

\subsubsection{Embedding inversion attacks}

Text embedding models map text to vectors. They capture semantics and other important features of the input text. Embedding inversion attacks target the front end of LLMs. They aim to recover the original text input \(x^*\) from its text embedding \(\varphi(x^*)\) by exploiting access to the embedding function \(\varphi\) of the LLM. 

Typically, embedding inversion aims to recover the original text \( x \) from its embedding \( e = \varphi(x) \) by maximizing the cosine similarity between the embedding of a candidate text \( \hat{x} \) and \( e \). 

In white-box scenarios, the attacker optimizes the difference between the target embedding and candidate inputs using continuous relaxation methods for efficient gradient-based inversion. In black-box settings, the attacker trains an inversion model using auxiliary data to predict the set of words in \(x^*\), employing techniques like multi-label classification or multi-set prediction to recover the original words without their exact order \cite{10.1145/3372297.3417270}. 
Assuming that collisions are rare, the attacker queries the model with candidate texts until he finds the text \( \hat{x} \) whose embedding closely matches \( e \) \cite{morris2023textembeddingsrevealalmost}.

Malicious actors with access to the vector corrsponding to some text embedding, as well as the API of an Embedding as a Service (EaaS) platform, can train an external model to approximate an inversion function and thus reconstruct the original text from its representation in the form of text embedding. 

There are also multilingual embedding inversion attacks that target multilingual LLMs \cite{chen2024textembeddinginversionsecurity}.
\section{Defenses against attacks on language models}
\label{chapter:resultados}

As we discussed in the previous Section, LLMs are vulnerable to several types of attacks. Various defense strategies have been developed to mitigate these risks and protect both sensitive training data and the integrity of the models. As in any other ML system, these defenses try to balance the need to maintain the performance and functionality of LLMs with the need to ensure their security and privacy \cite{varshney2023artdefendingsystematicevaluation}. 

In this Section, we survey existing defenses against adversarial attacks on LLMs. In general, any ML system can suffer both causative and exploratory attacks \cite{10.1145/1128817.1128824}. Regularization techniques can be useful to constrain what the ML system learns and, therefore, to increase its robustness against causative attacks. For exploratory attacks, we cannot expect the learning algorithm to be secret, yet disinformation to confuse attackers and information hiding (e.g. preventing the adversary from discovering the ML system hyperparameters) might raise the bar for attackers to succeed. It should also be noted that targeted attacks are more sensitive than indiscriminate attacks. For the latter, randomization can be effective: the adversary obtains imperfect feedback from the ML system, so more work is required for the attack to be successful, at the potential cost of decreasing the ML system performance. 

Given that LLM uses are varied and complex, and considering how learning algorithms use and interpret training data, specific solutions are often required. Customized defenses must be based on a deep understanding of the LLM training and inference process, and will vary depending on the particular context and the potential attack vectors \cite{tete2024threatmodellingriskanalysis}. 

Defense mechanisms against LLM attacks can be divided into two main categories: prevention and detection. 

\subsection{Prevention-based defenses}

Preventive defenses focus on redesigning the instruction prompt or preprocessing the data so that the model can perform its task even when the data has been compromised \cite{liu2024exploringvulnerabilitiesprotectionslarge}. These strategies focus on restructuring inputs and enhancing model robustness through mechanisms that isolate malicious elements, refine data representation, and fortify alignment with intended behaviors. Various approaches have been developed to address vulnerabilities in LLMs \cite{liu2024formalizingbenchmarkingpromptinjection}.

\subsubsection{Paraphrasing}

Data-driven techniques for generating paraphrases are useful for modifying inputs while preserving the original meaning, particularly in the context of adversarial defense strategies \cite{10.1162/coli_a_00002}. These methods have been explored extensively, leveraging generative frameworks to encode and transform text into alternative representations while maintaining semantic fidelity \cite{meng2017magnettwoprongeddefenseadversarial}. Recent advances have extended the application of paraphrasing techniques to LLMs for addressing adversarial prompts. 
Paraphrasing, as a preprocessing step, transforms potentially harmful inputs into benign harmless ones.
The generative model would accurately preserve natural instructions but would not reproduce an adversarial sequence of tokens accurately enough to maintain adversarial behavior \cite{jain2023baselinedefensesadversarialattacks}.
By rephrasing the adversarial instructions, this mechanism disrupts the structured patterns that adversaries rely on to exploit model vulnerabilities, while retaining the core intent of legitimate instructions. Empirical evaluations have shown that paraphrasing can effectively neutralize adversarial behavior in various settings, albeit with some trade-offs in accuracy and performance when applied to benign inputs flagged erroneously by adversarial detectors \cite{kirchenbauer2024reliabilitywatermarkslargelanguage}.

\subsubsection{Retokenization}

Retokenization consists of breaking up tokens in a prompt and representing them using multiple smaller tokens. As happened with paraphrasing, retokenization disrupts suspicious adversarial prompts without significantly degrading or altering the behavior of the model in case the prompt is benign. 

BPE-dropout \cite{provilkov-etal-2020-bpe} can be used to perform retokenization. High-frequency words are kept intact in the text, while rare words are broken down into multiple tokens. This is achieved by randomly dropping a percentage of BPE (Byte Pair Encoding) merges during text tokenization, which results in a random tokenization with more tokens than a standard representation \cite{jain2023baselinedefensesadversarialattacks}.

Retokenization is effective against jailbreaking and prompt injection attacks, since it alters the structure of the input text so that malicious instructions become less effective.

\subsubsection{Delimiters (e.g. spotlighting \& prompt data isolation)}

Delimiters can be introduced to separate data from instructions within the input prompt. Delimiters help the model focus on the intended instructions while ignoring compromised content \cite{xu2024llmsabstractionreasoningcorpus}.
As in query parameterization for preventing SQL injection attacks in relational databases, the use of specific delimiters, such as guillemets (<< >>) or triple single quotes (\rq \rq \rq), can help LLMs interpret the text between them as data, rather than valid prompt instructions, thus protecting against malicious prompt injections. By isolating \cite{liu2024automaticuniversalpromptinjection} or spotlighting \cite{hines2024defendingindirectpromptinjection} data with the help of delimiters, the model can focus on the intended task without executing harmful instructions embedded within the injected user-provided data. When delimiters are successful, they prevent prompt injection attacks from succeeding.

\subsubsection{Sandwich prevention}

Sandwich prevention involves constructing a carefully framed prompt where safe contextual instructions are placed both before and after the potentially harmful or adversarial input. By surrounding the compromised data with these "safety buffers," the model is guided to realign its behavior with the intended task and to resist adversarial manipulations embedded within the user-provided prompt. This sandwich is intended to reinforce the safety mechanisms of the LLM, ensuring that it adheres to the target task and reducing the likelihood of generating harmful or inappropriate responses. Sandwich prevention addresses weaknesses such as sensitivity to token length, multilingual vulnerabilities, and contextual confusion by systematically employing safety-focused prompts \cite{Upadhayay_2024}.

\subsubsection{Instructional prevention}

Instructional prevention focuses on explicitly reinforcing the importance of following the original prompt and not paying attention to possible external manipulations. This acts as a clear directive for the LLM to ignore any adversarial instructions contained in the compromised data. It consists of just instructing the LLM to ignore any instructions contained in the user-provided text and to adhere exclusively to the application-specific instruction prompt \cite{instruction_defense_2023}, a weaker form of spotlighting \cite{hines2024defendingindirectpromptinjection}.

\subsubsection{Embedding purification}

Embedding purification targets potential backdoors in word embeddings, detecting differences between pre-trained weights and compromised weights. It refines these embeddings to ensure they do not contain malicious triggers, thereby enhancing the model security and integrity.

For each word \( w_i \), let \( f_i \) represent its frequency in a large-scale clean corpus, and \( f_i' \) its frequency in a poisoned dataset. $\delta_i = E_{i,\text{backdoored}} - E_{i,\text{pretrained}}$ is the embedding difference vector of dimension \( n \). Under the assumption that  \( f_i' \approx C f_i \) for clean words, where \( C \) is a constant, $\|\delta_i\|_2 \propto \log(f_i)$ for clean words, whereas for trigger words $\frac{\|\delta_k\|_2}{\log(f_k)} \gg \frac{\|\delta_i\|_2}{\log(f_i)}$. This disproportionate discrepancy for trigger words allows the identification of compromised embeddings. Embedding purification \cite{zhang2022finemixingmitigatingbackdoorsfinetuned} then resets the embeddings of such words (e.g., the top 200 ranked by $\frac{\|\delta_i\|_2}{\log(\max(f_i, 20))})$
to their pre-trained values while preserving other embeddings.

\subsubsection{SmoothLLM}

SmoothLLM \cite{robey2024smoothllmdefendinglargelanguage} was designed against jailbreaking attacks, which aim to manipulate the model into generating inappropriate or undesirable content. SmoothLLM is based on a randomized smoothing technique that  introduces small random perturbations into multiple copies of the original input and then aggregates the responses generated by each perturbed copy. This allows for the detection and mitigation of adversarial inputs by identifying unusual patterns in the aggregated responses. SmoothLLM may, however, result in a slight decrease in model performance, but this reduction is moderate and can be adjusted by selecting the appropriate hyperparameters. This e method does not require retraining the underlying model and is compatible with both black-box and white-box models.

Let a prompt \( P \) and a distribution \( P_q(P) \) over perturbed copies of \( P \) be given. Let \( \gamma \in [0,1] \) and \( Q_1, \dots, Q_N \) be drawn i.i.d. from \( P_q(P) \). Let us now define \( V \) to be the majority vote of the jailbreaking  function $JB$ across these perturbed prompts with respect to the margin \( \gamma \), i.e., $V = \mathbb{I} \left( \frac{1}{N} \sum_{j=1}^{N} \left( \text{JB} \circ \text{LLM} \right) (Q_j) > \gamma \right)$. The $JB$ function is a binary-valued function that checks whether a response generated by an LLM constitutes a jailbreak. Then, \( \text{\sc SmoothLLM} \) is defined as $\text{\sc SmoothLLM}(P) = \text{LLM}(Q)$, where \( Q \) is any of the sampled prompts that agrees with the majority, i.e., $(\text{JB} \circ \text{LLM})(Q) = V$.

\subsubsection{Dimensional masking}

Dimensional masking \cite{chen2024textembeddinginversionsecurity} is a defense technique against embedding inversion attacks. This method tries to protect the information encoded in the embeddings by masking part of their content to hinder attackers from reconstructing the original text from the embeddings. In dimensional masking, the embeddings generated by a language model are modified by adding a language identification vector, i.e. is a vector that represents the language of the original text. By masking the first dimension of the embedding with this language identification vector, it becomes more difficult for an attacker to invert the embeddings and retrieve the original text. 

For an input text \( x \), the masked embedding function \( \phi_{\text{masking}} \) is defined as $\phi_{\text{masking}}(x) = \text{vec}\left([\text{id}_t, \text{vec}(\phi_i(x))_{1 \leq i \leq n}]\right)$, where 
$\phi(x) = \text{vec}(\phi_i(x))_{0 \leq i \leq n}$ is the original embedding, 
\( n \in \mathbb{N} \) is the dimensionality of \( \phi(x) \), 
\( \text{id}_t \in \mathbb{R} \) is a unique identifier encoding the target language \( l_t \), 
and \( \text{vec}(\cdot) \) denotes vectorization. This simple approach effectively reduces the success rate of embedding inversion attacks while fully preserving utility in retrieval tasks \cite{chen2024textembeddinginversionsecurity}. 

\subsubsection{Differential privacy}

Differential privacy (DP) originates from the idea of adding calibrated noise to the results of sensitive queries so that the presence or absence of an individual in the dataset does not significantly affect their outcome \cite{10.1007/978-3-540-28628-8_32}. Building on earlier work, differential privacy was formally defined by introducing a rigorous framework that uses the sensitivity of a query to calibrate the amount of noise to be added. DP ensures that the output remains statistically consistent while providing strong guarantees that individual contributions cannot be distinguished, even by adversaries with additional knowledge \cite{10.1007/11681878_14}.

In DP, randomized algorithms add calibrated noise to ensure that the probability of any output remains nearly unchanged, regardless of the presence or absence of any individual in the dataset, thus ensuring that individual privacy. 
A randomized function \( K \) provides \( \epsilon \)-differential privacy \cite{10.1007/11787006_1} if for all datasets \( D_1 \) and \( D_2 \) that differ in at most one element, and for all subsets \( S \subseteq \text{Range}(K) \), $\Pr[K(D_1) \in S] \leq e^{\epsilon} \cdot \Pr[K(D_2) \in S]$.

Extending the principles of differential privacy, techniques were developed to enable its application in deep learning, specifically through the Differentially-Private Stochastic Gradient Descent (DP-SGD) algorithm \cite{Abadi_2016}. DP-SGD balances privacy guarantees with model utility by calibrating noise to the sensitivity of gradient computations. An algorithm \( A \) satisfies \((\epsilon, \delta)\)-differential privacy if, for any pair of adjacent databases \( D \) and \( D' \) (where only one entry differs), and for any set of possible outcomes \( S \subseteq \text{Range}(A) \), the following inequality holds:
$\Pr[A(D) \in S] \leq e^{\epsilon}  \cdot \Pr[A(D') \in S] + \delta$.
Here, \( \epsilon \) quantifies the gap between the probabilities of obtaining a given result when a particular data point is present or absent, whereas \( \delta \) is a parameter that controls the likelihood of differential privacy being violated.

DP protects the privacy of the individuals who are part of the training dataset \cite{kurakin2024harnessinglargelanguagemodelsgenerate} ensuring that their presence or absence in the training set does not significantly affect the model output \cite{janryd2024preventing}. DP is highly effective in mitigating membership inference attacks and provides a mathematical guarantee on privacy, so that no sensitive information can be deduced for individual data instances in the training set \cite{info15110697}.

\subsubsection{Fine-tuning}

Fine-tuning was one of the earliest defenses against backdoor attacks. Initially, researchers explored basic fine-tuning techniques that retrain the model with clean data to mitigate backdoor effects. This method was first introduced as a way to adjust the weights of a model that had been compromised by malicious inputs to reset the desired model behavior \cite{gu2019badnetsidentifyingvulnerabilitiesmachine}. 
Fine-tuning methods evolved to account for the specific characteristics of backdoor attacks. A variation of this approach, known as fine-mixing, was proposed to combine both poisoned and clean data in a more sophisticated manner. Fine-mixing first merges the contaminated model weights with the pre-trained model weights, and then refines the combination using a limited set of clean data, allowing for better preservation of the model overall functionality while removing the backdoor effects \cite{liu2018trojaning}. 
Fine-tuning continues to be an essential white-box defense against adversarial attacks \cite{liu2024exploringvulnerabilitiesprotectionslarge}.

\subsection{Detection-based defenses}

Detection-based defenses aim to identify when the prompt has been compromised, using monitoring algorithms and constant auditing of the model responses to detect suspicious patterns or anomalous behavior \cite{liu2024exploringvulnerabilitiesprotectionslarge}.

\subsubsection{Perplexity-based detection}

Perplexity was originally proposed in the context of speech recognition \cite{jelinek1977perplexity} and is commonly used in NLP for measuring the quality of a language model. In information theory, perplexity is a measure of uncertainty in the value of a sample from a discrete probability distribution. The larger the perplexity, the less likely it is that an observer can guess the value which will be drawn from the distribution.

Perplexity per token is mathematically defined as 
$PPL(x) = \exp\left( -\frac{1}{t} \sum_{i=1}^{t} \log p(x_i \mid x_{<i}) \right)$, 
where \(x = (x_1, x_2, \dots, x_t)\) is a sequence of \(t\) tokens, and \(p(x_i \mid x_{<i})\) is the conditional probability of the token \(x_i\) given all preceding tokens \(x_{<i}\). A lower perplexity indicates that the sequence is more "natural" according to the model. Perplexity (per token) is an information theoretic measure that evaluates the similarity of proposed model to the original distribution and it can be interpreted as the exponentiated cross entropy.

Perplexity-based detection (PPL) \cite{alon2023detectinglanguagemodelattacks} is based on the observation that queries with adversarial suffixes have exceedingly
high perplexity values.  Since injecting adversarial instructions or data into a text often increases its perplexity, a simple detector can be designed: when the perplexity of the input text exceeds a predefined threshold, it is considered to be compromised.


A classifier can also be trained to differentiate between adversarial and non-adversarial inputs by considering perplexity and token sequence length (another indicator of potentially adversarial inputs). Taking both signals into account (perplexity and length), the resulting classifier significantly outperforms simple perplexity thresholding.

Another variant, windowed perplexity detection \cite{jain2023baselinedefensesadversarialattacks}, divides the input text into contiguous windows and calculates the perplexity of each window. If any window exceeds the threshold, the input is considered to be compromised.

While machine-generated adversarial attacks exhibit high perplexity and long token sequences, human-crafted adversarial attacks tend to mimic normal text with low perplexity and lengths that are similar to benign inputs, making their detection more challenging.

\subsubsection{Naive LLM-based detection}

The LLM itself can be used to detect compromised data. The LLM is queried with a specific instruction about the content of the input text in order to determine whether it is compromised or clean. If the LLM indicates that the provided text is compromised, it is acted upon accordingly. The defense relies on the LLM ability to identify whether a prompt includes malicious or unauthorized instructions \cite{Wong:EECS-2024-135}.

\subsubsection{Response-based detection}

Response-based detection relies on the LLM prior knowledge of the expected response for a specific task. In contrast to the naive LLM-based detection above, response-based defenses first generate a response before evaluating whether the response is harmful. If the response generated by the LLM does not match valid responses expected for the target task, the input prompt is considered to be compromised. A key limitation of this defense mechanism is that it fails when both the injected task and the target task are of the same kind \cite{zeng2024autodefensemultiagentllmdefense}.

As in naive LLM-based detection, response-based detection can leverage the intrinsic capabilities of LLMs to evaluate the response \cite{phute2024llmselfdefenseself} rather that the input prompt. Backtranslation \cite{wang2024defendingllmsjailbreakingattacks} is another potential strategy: given the initial LLM-generated response, prompts the LLM to infer an input prompt that can lead to that response (the backtranslated prompt), which tends to reveal the actual intent of the original prompt, and run the LLM again on the backtranslated prompt (i.e. the original prompt is rejected if the model refuses the backtranslated prompt).
A third alternative consists of making the LLM aware of potential harm by asking it to repeat its response \cite{zhang2024pardenrepeatthatdefending}: when the system is unable to repeat the LLM output generated from a malicious user input, the BLEU score between the original LLM output and repetition falls below a similarity threshold (a hyper-parameter of the detection method); then, the user input classified as malicious and the repeated output is returned to the output instead of the original output.

Another approach for response-based detection is based on the use of classifiers, such as Llama Guard \cite{inan2023llamaguardllmbasedinputoutput} and Self-Guard \cite{wang2024selfguardempowerllmsafeguard}, which are applied to categorize prompt-response pairs into safe or unsafe categories, enhancing robustness against adversarial prompts.

\subsubsection{Perturbation}

Perturbing the LLM input can also help us detect the presence of malicious inputs. Inserting a perturbation including rare words into the input text provokes different effects depending on the particular situation. When adding the perturbation to a clean input, the model output probability of the target class drops. However, when adding this rare word perturbation to a poisoned sample, the confidence of the target class does not change too much, since the attacker’s goal is to make the trigger work in a wide range of situations.

That difference between clean and malicious inputs can be used to create an efficient online defense mechanism based on robustness-aware perturbations (RAP) \cite{ yang2021raprobustnessawareperturbationsdefending}. RAP exploits the gap of robustness between clean and contaminated samples, so they can be distinguished. RAP provides an effective defense mechanism against backdoor attacks on LLMs at inference time. 

\subsubsection{Masking-differential prompting}

Masking-differential prompting (MDP) \cite{xi2023defendingpretrainedlanguagemodels} is a defense mechanism designed to mitigate backdoor attacks in pre-trained language models (PLMs) operating as few-shot learners. MDP leverages the gap between the masking-sensitivity of poisoned and clean samples: with reference to the limited few-shot data as distributional anchors, MDP compares the representations of given samples under varying masking and identifies poisoned samples as the ones with significant variations. 

MDP capitalizes on the higher sensitivity of poisoned samples to random masking when compared to clean samples. This method measures the representational change of a sample before and after random masking, allowing the identification of poisoned data. Specifically, the model sensitivity to masking is quantified by $\tau(X_{\text{test}}) = \Delta(d(X_{\text{test}}), d(\hat{X}_{\text{test}}))$, where
\( d(X_{\text{test}}) \) denotes the feature vector of a test sample \( X_{\text{test}} \), and \( \hat{X}_{\text{test}} \) is the same sample after random masking. The function \( \Delta \) measures the distance between the original and masked representations, and the variation \( \tau(X_{\text{test}}) \) is used to classify a sample as poisoned when the variation \( \tau(X_{\text{test}}) \) exceeds a defined threshold.


\subsubsection{Anomaly detection}

Anomalies, also called outliers in Statistics, stand out from the rest of data. They deviate so much from other observations that they lead to suspicion that they were generated by a different mechanism \cite{hawkins1980identification}. 

Defense mechanisms based on anomaly detection rely on identifying poisoned examples in the training dataset by detecting outliers or anomalies. Anomaly detection can be performed using features such as text embeddings or the perplexity of texts. 

A well-known regularization technique such as early stopping can be used to reduce the number of training epochs as a simple mechanism that limits the impact of poisoning (apart from overfitting), achieving a moderate defense against poisoning at the cost of
some prediction accuracy. Methods designed for the identification and removal of poisonous examples from data are much more effective. Multiple strategies based on anomaly detection are possible \cite{wallace2021concealeddatapoisoningattacks}. For instance, poisoned examples often contain phrases that are not fluent English, so perplexity can be used to identify non-fluent sentences. Text embeddings can also be useful for locating poisoned examples: even when poisoned examples have no lexical overlap with trigger phrases, their embeddings might appear suspiciously similar.
Extending this approach, recent advancements leverage semantic anomaly detection techniques \cite{Elhafsi2023}: LLMs themselves can identify inconsistencies in sentence-level semantics and detect inputs that deviate significantly from expected patterns. 


\section{Coverage of defenses against attacks on LLMs}

Large language models, such as GPT, Gemini, Llama, Grok, or Claude, face a variety of security risks arising from their ability to process and generate text based on large and diverse training databases \cite{das2024securityprivacychallengeslarge}. Malicious actors can exploit LLM vulnerabilities by performing a wide range of attacks on LLM-based systems. Given the increased use of these models in sensitive applications, it is crucial to understand how different kinds of attacks can be prevented and/or mitigated with the help of existing defense mechanisms.

\newcommand{\xtarget}{\ding{55}} 
\begin{table*}
  \caption{Attacks on LLMs, including the acronyms used in Table ~\ref{tab:defenses} and whether they taget model integrity and/or data privacy.}
  \label{tab:attacks}
  \begin{tabular}{lclcc}
    \toprule
          &         &        & Model     & Data \\
    Phase & Acronym & Attack & integrity & privacy \\
    \midrule
    During training... 
    & \texttt BA & Backdoor attack & \xtarget & \\
    & \texttt DP & Data poisoning & \xtarget & \\
    & \texttt GL & Gradient leakage &  & \xtarget \\
    \midrule
    On trained models... 
    & \texttt AI & Adversarial input & \xtarget & \\
    & \texttt JB & Jailbreaking & \xtarget & \xtarget \\
    & \texttt PI & Prompt injection & \xtarget & \xtarget \\
    & \texttt MI & Model inversion &  & \xtarget \\
    & \texttt DE & Data extraction &  & \xtarget \\
    & \texttt MIA & Membership inference attack &  & \xtarget \\
    & \texttt EI & Embedding inversion &  & \xtarget \\
    \bottomrule
  \end{tabular}
\end{table*}


\begin{sidewaystable}
    \centering
    \newcommand{\xmark}{\ding{55}}%

\begin{tabular}{ll|c|c|c|c|c|c|c|c|c|c|l}
\cline{3-12}
& & \multicolumn{3}{c|}{\textbf{\begin{tabular}[c]{@{}c@{}}Attacks\\during\\training\end{tabular}}} 
  & \multicolumn{7}{c|}{\textbf{\begin{tabular}[c]{@{}c@{}}Attacks\\on trained\\models\end{tabular}}}                                                               & \\ \hline
\textbf{Category} & \textbf{Defense} & \multicolumn{1}{c|}{BA}       & \multicolumn{1}{c|}{DP}       & \multicolumn{1}{c|}{GL}       & \multicolumn{1}{c|}{AI} & \multicolumn{1}{l|}{JB} & \multicolumn{1}{l|}{PI} & \multicolumn{1}{c|}{MI} & \multicolumn{1}{c|}{DE} & \multicolumn{1}{c|}{MIA} & EI & \textbf{Effectiveness} \\ \hline
\multirow{10}{*}{\textbf{Prevention}} 
& Paraphrasing   &          &          &          & \checkmark  & \checkmark  & \checkmark  &    &    &     &    & Moderate \\ \cline{2-13} 
& Retokenization &          &          &          & \checkmark  & \checkmark  & \checkmark  &    &    &     &    & High   \\ \cline{2-13} 
& Delimiters     &          &          &          & \checkmark  & \checkmark  & \checkmark  &    &    &     &    & Moderate  \\ \cline{2-13} 
& Sandwich prevention &          &          &          &    &    & \checkmark  &    &    &     &    & Moderate   \\ \cline{2-13} 
& Instructional prevention  &          &          &          &    &    & \checkmark  &    &    &     &    & Moderate \\ \cline{2-13} 
& Embedding purification & \checkmark        &          &          &    &    &    &    &    &     &    & Moderate  \\ \cline{2-13} 
& SmoothLLM   &          &          &          &    & \checkmark  &    &    &    &     &    & High  \\ \cline{2-13} 
& Dimensional masking &          &          &          &    &    &    &    &    &     & \checkmark  & High   \\ \cline{2-13} 
& Differential privacy &          &          & $\sim$ &    &    &    &    &    & \checkmark   &    & Low (GL) / High (MIA) \\ \cline{2-13}
& Fine-tuning & \checkmark        &          &          &    &    &    &    &    &     &    & High  \\  \hline
\multirow{6}{*}{\textbf{Detection}}   
& Perplexity-based detection  &          &          &          &    & \checkmark  & \checkmark  &    & \checkmark  &     &    & Moderate \\ \cline{2-13} 
& Naive LLM-based detection &          &          &          &    &    & \checkmark  &    &    &     &    & Moderate  \\ \cline{2-13} 
& Response-based detection &          &          &          &    &    & \checkmark  &    &    &     &    & Moderate \\ \cline{2-13} 
& Perturbation  & \checkmark        &          &          &    &    &    &    &    &     &    & Moderate  \\ \cline{2-13} 
& Masking-differential prompting & \checkmark        &          &          &    &    &    &    &    &     &    & Moderate \\ \cline{2-13} 
& Anomaly detection &          & \checkmark        &          &    &    &    &    &    &     &    & High    \\ \hline
& \textbf{Without known defenses}  &          &          &          &    &    &    & \xmark  &    &     &    & N/A \\ \cline{2-13} 
\end{tabular}

    \vspace{1em}
    \caption{Available defenses for different kinds of attacks on LLMs and their effectiveness.}
    \label{tab:defenses}
\end{sidewaystable}

Table \ref{tab:attacks} lists the kinds of attacks that exploit LLM vulnerabilities in different stages of their life cycle, as well as their ultimate target, which might be compromising model integrity or getting access to private data.

Table \ref{tab:defenses} shows how existing defense mechanisms address the vulnerabilities exploited by different kinds of attacks on LLMs, providing an eagled-eyed view view of the LLM security landscape. To the best of our knowledge, it summarizes the current state of the art. Apart from indicating which defense mechanisms and techniques are effective against different kinds of attacks, it also estimates the protection effectiveness they provide. A quick recap is now in order:

\begin{itemize}

\item
For backdoor attacks, embedding purification is moderately effective because, although cleaning the embedded representations can decrease the potential attack surface area, it can never guarantee the prevention of all possible backdoors, especially the more sophisticated ones. Perturbation is also moderately effective, as masking-differential prompting, because backdoor attacks can be designed to be robust against perturbations and masking. Fine-tuning and fine-mixing could, in principle, be highly effective to identify and remove backdoors. 

\item
Against data poisoning attacks, anomaly detection can be highly effective, in the same sense that fine-tuning was effective for backdoor attacks. In principle, given sufficiently advanced anomaly detection techniques, the attacker might find it difficult to evade detection.

\item
The effectiveness of differential privacy against gradient leakage attacks has proven in practice to be low for several reasons.
First, differential privacy relies on adding noise to gradients to hide the information they reveal about training data. While this technique can be effective in protecting individual data, in the context of LLMs, the additional noise may not be sufficient to mitigate more sophisticated attacks that exploit patterns and correlations underlying gradients on a large scale. 
Furthermore, implementing differential privacy in LLMs requires tuning parameters such as the noise level, which often represents a critical trade-off between privacy and model utility. A high noise level to ensure privacy can significantly degrade the model performance.

\item
In adversarial input attacks, paraphrasing and retokenization techniques demonstrate moderate and high levels of effectiveness, respectively. The effectiveness of the use of delimiters is also moderate.

\begin{itemize}

\item Paraphrasing is moderately effective because it can mislead or disrupt adversarial inputs by changing superficial input text features (i.e. its wording). However, its effectiveness may be limited, as those transformations alone might not be sufficient to prevent a well-crafted adversarial input from succeeding. 

\item Conversely, retokenization is highly effective because changing token segmentation alters how the model interprets the input at a more fundamental level, making it more difficult for attackers to predict how their malicious inputs will be processed by the model and significantly reducing the efficacy of such attacks.

\item Delimiters are also moderately effective in defending against adversarial inputs. Spotlighting and prompt data isolation provide some protection by separating different parts of the input, yet a clever attacker might circumvent such defenses the same way SQL-injection attacks can be performed on SQL databases. 

\end{itemize}

\item
Against jailbreaking attacks, paraphrasing, retokenization, and the use of delimiters demonstrate varying levels of effectiveness, as each approach prepares the input differently. Detection-based techniques also exhibit different degrees of success against jailbreaking.

\begin{itemize}

\item Paraphrasing is moderately effective, as it reformulates inputs, potentially disrupting some jailbreaking attempts by altering the structure and content of the input. 

\item Retokenization can be highly effective against jailbreaking attacks, 
fundamentally changing how the model interprets the input and making jailbreak attacks considerably harder to execute successfully. 

\item Delimiters are only moderately effective because, even when they try to enforce a clear, rigid structure within the input, they cannot guarantee it in the same way a parameterized SQL query is interpreted by a RDBMS to prevent injection attacks (the LLM is still there as the final interpreter of the input).

\item Perplexity-based detection (PPL) shows moderate effectiveness against jailbreaking attacks, since it can spot some jailbreaking attempts but is far from infallible.

\item SmoothLLM, on the other hand, can be highly effective,
enhancing the model resilience to malicious inputs and limiting the attackers' ability to jailbreak the model (as anomaly detection against data poisoning).

\end{itemize}

\item
For prompt injection attacks, multiple defenses have been devised due to the high prevalence of these attacks that compromise the integrity of LLMs. 

\begin{itemize}

\item Retokenization, spotlighting, and prompt data isolation address input manipulation from different angles. Retokenization makes it difficult for attackers to manipulate the input, hence its effectiveness can be high. Delimiter-based techniques, such as spotlighting and prompt data isolation, have inherent limitations that make them vulnerable to sophisticated attacks.

\item Paraphrasing, sandwich prevention, and instructional prevention try to frame user-provided inputs in different ways. They are moderately effective against prompt injection attacks, as attackers can find sophisticated ways to evade these defense mechanisms.

\item Perplexity-based (PPL), naive LLM-based, and response-based detection monitor the behavior of the model when faced with suspicious inputs. They are also moderately effective. A canny attacker might find creative ways to surpass those defenses.

\end{itemize}

\item 
At the time of this writing, no specific defense mechanisms are known against model inversion attacks.

\item
Data extraction attacks can be mitigated by perplexity-based detection (PPL). Its effectiveness is limited (i.e. moderate) because attackers can craft their input in a way that keeps perplexity within the range expected by the model, thus avoiding detection.

\item
Against membership inference attacks (MIA), differential privacy (DP) is the way to go. DP is highly effective defense due to its fundamental ability to protect individual privacy in the context of machine learning models. 

\item
Finally, dimensional masking defense is highly effective against embedding inversion attacks. First, this technique causes a misalignment in the interpretation of the embedding, drastically reducing the attackers' ability to recover the original text. Second, it does so without compromising the LLM performance.

\end{itemize}

\section{Conclusion}


In this paper, we have explored the risks associated to the use of large language models (LLMs), which affect both the integrity of LLM-based systems and the privacy of the data LLMs are trained with.

Known LLM vulnerabilities have been exploited by different kinds of attacks, which can act both when the LLM is being trained (i.e. during training) and also when it is deployed in practice (i.e. during inference). Our survey has analyzed LLM attack vectors and the defensive mechanisms that have been devised to prevent, mitigate, or detect those attacks.

Even though it should be noted that security is always relative, our study of the coverage of defense mechanisms against attacks on LLM-based systems has highlighted areas that require additional attention to ensure the reliable use of LLMs in sensitive applications. Only a handful of defenses are highly effective, in the sense that the can be made as sophisticated as the sophistication of the attacker might require them to be. Most existing defense mechanisms and countermeasures, unfortunately, present inherent limitations that can be exploited by informed and sufficiently-skilled malicious actors.



\bibliographystyle{ACM-Reference-Format}
\bibliography{references}

\end{document}